\documentclass[twocolumn,showpacs,preprintnumbers,amsmath,amssymb]{revtex4}
\usepackage{CJK}
\usepackage{graphicx}
\usepackage{dcolumn}
\usepackage{bm}
\usepackage{color}

\begin{document}
\preprint{APS/123-QED}

\title{ Universality for Moving Stripes: A Hydrodynamic Theory of Polar Active Smectics}
\author{Leiming Chen}
\address{College of Science, The China University of Mining and Technology, Xuzhou Jiangsu, 200116, P. R. China}
\author{John Toner}
\affiliation{Department of Physics and Institute of Theoretical
Science, University of Oregon, Eugene, OR 97403}
\date{\today}
\begin{abstract}
We present a hydrodynamic theory of polar active smectics, for systems both with and without number conservation. For the latter, we find quasi long-ranged smectic order in $d=2$ and long-ranged smectic order in $d=3$. In $d=2$ there is a Kosterlitz-Thouless type phase transition from the smectic phase to the ordered fluid phase driven by increasing the noise strength. For the number conserving case, we find that giant number fluctuations are greatly suppressed by the smectic order; that smectic order is long-ranged in $d=3$; and that nonlinear effects become important in $d=2$.
 \end{abstract} 
\maketitle

Active matter \cite{schweitzer2003} can exhibit a richer variety of ordered phases than for equilibrium systems, since  symmetry differences which have little or no effect in equilibrium can have radical effects on  active systems. For example, in equilibrium, systems with ``polar"  orientational order (e.g., ferromagnets) and those with {\it apolar}  orientational order (i.e., nematics)\cite{Tom's book} have identical scaling of their orientational 
 fluctuations.  In contrast, the active polar order  of a coherently moving flock\cite{ActiveMagnet} fluctuates far less than that  in active {\it apolar} orientationally ordered systems (i.e., ``active nematics"\cite{actnem}).


In this paper we formulate the hydrodynamic theory of {\it polar} active smectic systems,
by which we mean systems that spontaneously form uniformly spaced moving layers, and find that they differ considerably from their apolar (i.e., non-moving) analogs\cite{ApolarSmectic}. Examples of such  ``active polar smectics" include propagating waves in chemical reaction-diffusion  systems\cite{reac-dif}, and ``flocks"\cite{ActiveMagnet} of active particles forming uniformly spaced  parallel liquid-like layers (density waves).  Layers are  ubiquitous\cite{banding} as the order-disorder transition is approached in, e.g., the ``Vicsek model"\cite{Vicsek} of flocking,   although they may not reach a uniform steady state spacing.

We restrict ourselves here to ``active smectics $A$",  meaning phases  with the average particle  velocity along the mean layer normal.``Active smectics C"\cite{deGennes' book},  with other relative orientations of particle velocity and the layers, will be considered elsewhere\cite{Pawel}.
We consider two  cases: first,  with no conserved quantities, and second, with only  particle number conserved. In neither case is momentum conserved.


For the non-number conserving case (hereafter  the  ``Malthusian" case\cite{malt}), we find that the active polar  smectic phase is stable over a finite range of parameters; that is, it is  both hydrodynamically stable, and  robust against noise. In particular, we find  quasi-long-ranged smectic order in $d=2$, and long-ranged order in $d=3$.
In contrast,  thermal fluctuations  completely destabilize   the  {\it equilibrium}  smectic phase in $d=2$ \cite{2dsmectic},  and allow only quasi-long-ranged  equilibrium smectic order in $d=3$ \cite{Caille}.

More specifically, in $d=2$ 
\begin{eqnarray}
 \langle \psi^*(\vec{r},t)\psi(\vec{r}\,',t)\rangle\propto|\vec{r}-\vec{r}\,'|^{-\eta},
 \label{quasi}
\end{eqnarray}
where $\psi$ is the complex smectic order parameter, defined, as in equilibrium smectics \cite{deGennes' book}, via :
\begin{eqnarray}
\rho(\vec{r},t)\equiv\rho_0+\psi(\vec{r}\,,t)e^{iq_0z}+\rm{c.c.}~,
\label{psidef}
\end{eqnarray}
where $\rho$ is the number density (or, more generally, the spatially modulated scalar field in the problem, such as chemical concentration in reaction-diffusion systems), $\rho_0$ its mean, and $q_0\equiv2\pi/a$, with $a$ the distance between neighboring layers.  Here we've defined the average plane of the  layers as the $\perp$ plane, and the normal to this plane
as the $z$ axis.
Equations (\ref{quasi}) and  (\ref{psidef}) imply quasi-sharp Bragg peaks in light or X-ray scattering (or, equivalently, the numerical Fourier transform of density correlations); that is:
\begin{eqnarray}
 I_n(\vec{q})\propto\left<|\rho(\vec{q},t)|^2\right>\propto \left[\left(q_z-nq_0\right)^2+\gamma q_{\perp}^2\right]^{-2+n^2\eta\over 2},\label{Xray}
\end{eqnarray}
where $\gamma$ is an $O(1)$  constant, $n$ is an integer denoting the order of the  Bragg peak, and $\eta$ is non-universal (i.e., it varies from system to system).

For a finite system, this divergence is cut off for $|\vec{\delta q}|\sim 1/L$, where $L$ is the spatial linear extent of the system, and $\vec{\delta q}\equiv\vec{q}-nq_0\hat{z}$. This implies the $n$th peak will have a finite height which scales with $L$ like $L^{2-n^2\eta}$.


In $d=3$, the long-ranged nature of the smectic order implies sharp (i.e., $\delta$-function) Bragg peaks. In a finite system, the height of these peaks scales linearly with system volume $L^3$.

\begin{figure}
 \includegraphics[width=0.4\textwidth]{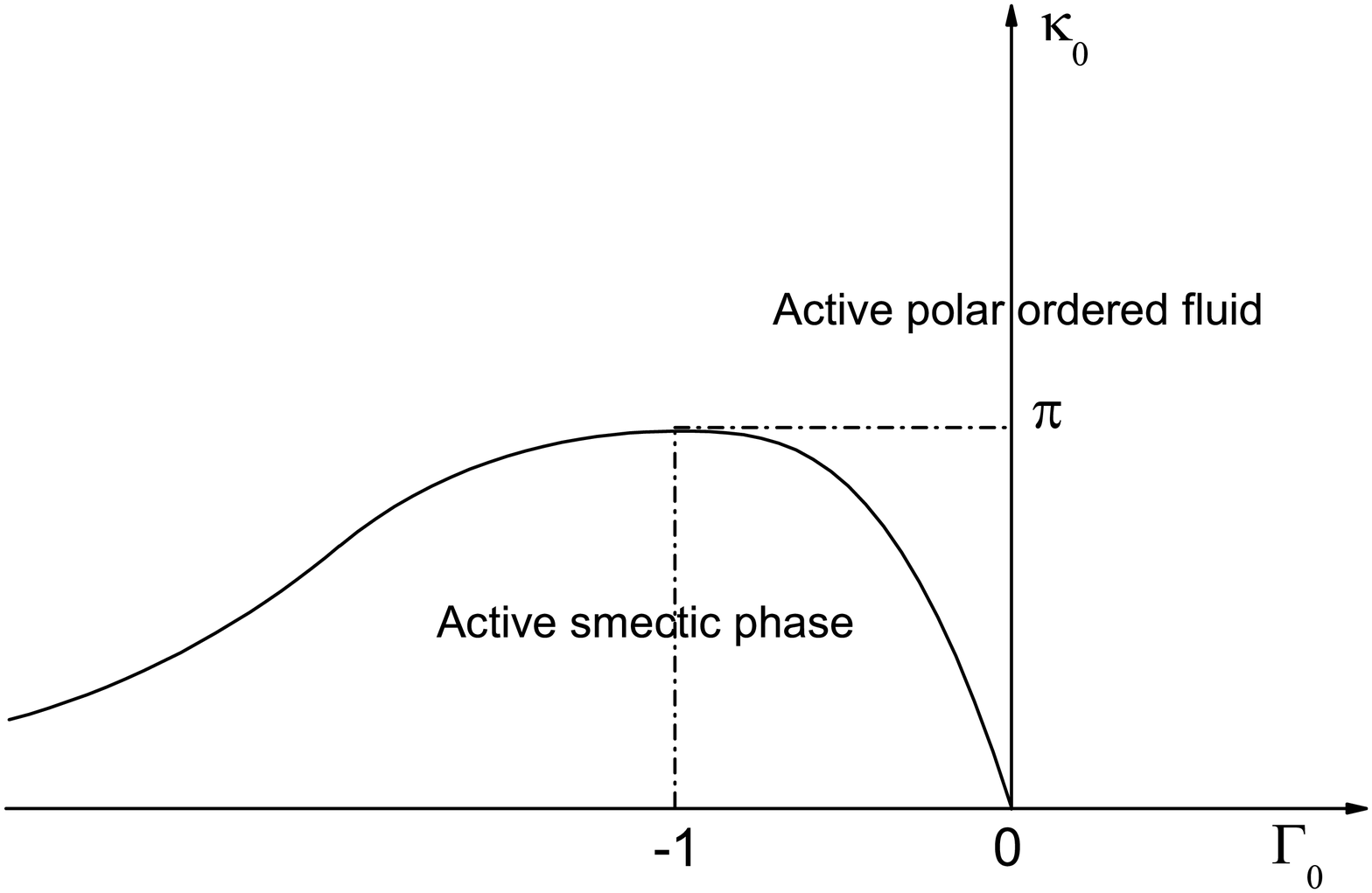}
 \caption{\label{fig: ActivePhase}Phase diagram in the $\kappa_0$-$\Gamma_0$ parameter space.}
\end{figure}

We predict that with increasing noise the active smectic phase undergoes
a dynamical phase transition
into  the fluid, polar ordered phase
treated in much prior work\cite{ActiveMagnet}. This transition is in the equilibrium XY universality class\cite{Tom's book}, which in $d=2$  is of  the the Kosterlitz-Thouless type\cite{KT}. The phase diagram in the parameter space of our model is  illustrated in Fig. \ref{fig: ActivePhase}.

In the case where the number of the particles is conserved, we find  long-ranged smectic order in $d=3$. In $d=2$,   the linearized version of the full hydrodynamic theory predicts quasi-long-ranged smectic order; however,  there are  marginal  non-linearities  which {\it may} invalidate this conclusion. We'll investigate this in a future publication\cite{future}.

In neither case  are there giant number fluctuations in $d=3$; nor are there are any  in  $d=2$ in the Malthusian case. The linearized hydrodynamic theory predicts none  in $d=2$ for the number conserving case either; but the aforementioned non-linear terms could  change this as well.

We'll now outline the derivation of these results, starting with the  case  without number conservation. Then the only important hydrodynamic variable is  the displacement $u(\vec{r}, t)$ of the layers  along $z$, 
which is proportional to the phase of the the smectic order parameter $\psi(\vec{r}, t)$: $\psi(\vec{r}, t)=\left|\psi_0\right|e^{-iq_0u(\vec{r}, t)}$\cite{deGennes' book, amplitude flucts nonhydro}.

Symmetry considerations (specifically, translation and rotation invariance) require that  $u$'s  equation of motion, to lowest order in  a gradient expansion, take the form:
\begin{eqnarray}
 \partial_t u&=&v_0-2\lambda_{\perp}\partial_z u + \left(\nu_{\perp}\nabla_{\perp}^2+\nu_z\partial_z^2\right)u\nonumber\\
 &&+ \lambda_z\left(\partial_z u\right)^2+ \lambda_{\perp}\mid\nabla_{\perp} u\mid^2 +f,
 \label{Original}
\end{eqnarray}
where 
$f$ is a Gaussian, zero-mean, white noise with variance $\left<f(\vec{r}, t)f(\vec{r}~', t')\right>=2\Delta\delta^d(\vec{r}-\vec{r}~')\delta(t- t')$.
Rotation invariance forces the coefficient of the $\partial_z u $ term to be exactly $-2$ times that of the $\mid\nabla_{\perp} u\mid^2$ term, because only the combination $\partial_z u-{1\over 2}\mid\nabla_{\perp} u\mid^2$ is unchanged by a uniform rotation of the smectic layers\cite{deGennes' book}.

The term with coefficient $\nu_\perp$ in (\ref{Original}) is forbidden by rotation-invariance of the free energy in
\textit{equilibrium}. It is, however,
permitted here\cite{active_tension, Ramaswamy2000} simply because rotation-invariance \textit{at the level of the
equation of motion}, which is all one can demand in an active system, does not
rule it out. Its physical content is that layer curvature produces a local
vectorial asymmetry which must modify the directed motion of the layers as this is
a driven system.

In contrast to {\it apolar} active smectics\cite{ApolarSmectic}, Eq. (\ref{Original}) is {\it not} invariant under the {\it simultaneous} transformation $u\to -u$, $z\to -z$, due to the lack (by definition) of up-down symmetry in {\it polar} smectics.

To simplify Eq. (\ref{Original}), we introduce another field variable
\begin{eqnarray}
u'=u-v_0t
\label{uprime}
\end{eqnarray}
and another coordinate system
 $t'=t,
 z'=z-2\lambda_{\perp} t,
 \vec{r}_\perp^{~\prime}= \vec{r}_\perp.
$ 
In terms of these 
Eq. (\ref{Original}) becomes
\begin{eqnarray}
\partial_{t'} u'&=& \nu_{\perp}\nabla_{\perp'}^2 u' + \nu_z\partial_{z'}^2 u' + \lambda_{\perp} |\vec{\nabla}_{\perp'}u'|^2+\lambda_z \left(\partial_{z'} u'\right)^2 \nonumber\\
&&+f,
\label{AnisoKPZ}
\end{eqnarray}
In Eq. (\ref{AnisoKPZ}), $\nu_{\perp, z}$ must be positive for the smectic state  to be dynamically stable. 
Either sign of  $\lambda_{\perp, z}$ can be stable; indeed, their signs need not be the same.

Eq. (\ref{AnisoKPZ}) has exactly the same form as the anisotropic KPZ equation\cite{AKPZ}. However, there is a crucial difference. The original KPZ equation\cite{KPZ} describes the hydrodynamics of crystal growth, and the hydrodynamic variable is $h$, the height of a $d$ dimensional surface. Clearly, states with different heights $h$ are always physically distinguishable. However, for smectics,  the state is periodic in $u'$ with period $a$, the spacing between neighboring smectic layers. This allows for the existence of topologically stable dislocations, which can unbind, thereby ``melting" (i.e., disordering) the smectic, in analogy to such ``dislocation mediated melting" in  a variety of translationally ordered equilibrium systems\cite{Undislocation, 2dsmectic}.
In $d=2$, this is the 
aforementioned
Kosterlitz-Thouless phase transition, which is absent in the anisotropic KPZ equation.

Simple power counting shows that the nonlinear terms in Eq. (\ref{AnisoKPZ}) are irrelevant in $d=3$; hence, the linear theory is valid.  
A  straightforward calculation then shows that $\langle |u'(\vec{q}, t)|^2\rangle\propto 1/q^2$ for all directions of wavevector $\vec{q}$. This in turn implies that  the real space fluctuation $\langle |u'(\vec{r}, t)|^2\rangle$ is finite as system size $L\to \infty$, which implies long-ranged smectic order\cite{foot1}.

In $d=2$ the nonlinear terms in Eq. (\ref{AnisoKPZ}) become marginal, and a dynamical renormalization group (RG) analysis is needed. This has already been done for the $d=2$  crystal growth problem \cite{AKPZ}; the   resulting RG recursion relations are:
\begin{eqnarray}
&&{d\nu_{\perp}\over d\ell}=\left[z-2+{g\over 32\pi}\left(1-\Gamma\right)\right]\nu_{\perp},\label{nuperp}\\
&&{d\lambda_{\perp,z}\over d\ell}=\left(\chi+z-2\right)\lambda_{\perp,z},\label{lambda}\\
&&{d\Delta\over d\ell}=\left[-2\chi+z-2+{g\over 64\pi}\left(3\Gamma^2+2\Gamma+3\right)\right]\Delta,\label{Delta}\\
&&{d\left(\nu_z/\nu_{\perp}\right)\over d\ell}=-{g\over 32\pi}{\nu_z\over\nu_{\perp}}\left(1-\Gamma^2\right),\label{nuz/nuperp}
\end{eqnarray}
where $\chi$ and $z$ are  the rescaling exponents of $u'$ and $t$, (i.e., $u'\to u'e^{\chi\ell}$, $t'\to t'e^{z\ell}$),  $\Gamma\equiv{\lambda_z\nu_{\perp}\over\nu_z\lambda_{\perp}},
 g\equiv{\Delta\lambda_{\perp}^2\over\nu_{\perp}^{5/2}\nu_z^{1/2}}$,
and we have chosen to rescale  lengths  isotropically. 
Eqns. (\ref{nuperp}-\ref{nuz/nuperp}) imply a set  of closed flow equations in $\Gamma$-$g$ space:
\begin{eqnarray}
 {d\Gamma \over d\ell}&=&{\Gamma g\over 32\pi}\left(1-\Gamma^2\right),\label{Gamma}\\
 {dg\over d\ell}&=&{g^2\over 32\pi}\left(\Gamma^2+4\Gamma-1\right)\label{g}.
\end{eqnarray}

\begin{figure}
 \includegraphics[width=0.4\textwidth]{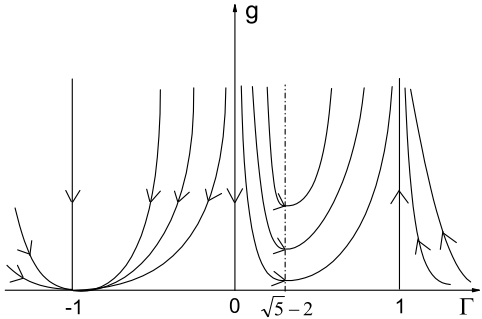}
 \caption{\label{fig: FGraph}The RG flow in the $\Gamma$-$g$ parameter space for active smectics in $d=2$. For $\Gamma<0$ and $g>0$, all flow lines go  to a stable fixed point $(-1,0)$; for $\Gamma>0$ and $g>0$, all flow lines go to infinity.}
\end{figure}

The ratio of these two equations yields a separable ODE
for $\Gamma$ as a function of $g$. Inserting the resultant solution for $\Gamma(g)$ into (\ref{g}) yields an integrable equation for $g(l)$; inserting that $g(l)$ into (\ref{Gamma}) yields a solvable
equation for $\Gamma(\ell)$. We thereby find
\begin{eqnarray}
 &&{\Gamma\over\left(1+\Gamma\right)^2}=\Gamma_0g_0\left(1-\Gamma_0\over {1+\Gamma_0}\right)^2\ell+{\Gamma_0\over\left(1+\Gamma_0\right)^2}, \label{Gammasol}\\
 &&g={\Gamma_0g_0\over\Gamma}\left(1-\Gamma_0\over 1-\Gamma\right)^2{\left(1+\Gamma\over 1+\Gamma_0\right)^2} \label{gsol},
\end{eqnarray}
where $g_0$ and $\Gamma_0$ denote respectively the ``bare" values of $g$ and $\Gamma$ (i.e., their values  at $\ell=0$). This solution implies a stable fixed point for $\Gamma_0<0$: as $\ell\to\infty$, $\Gamma(\ell)\to-1$, $1+\Gamma(\ell)\propto 1/\sqrt{\ell}$, and $g(\ell)\propto 1/\ell$. The RG flow loci in$\Gamma$-$g$ space are illustrated in Fig. \ref{fig: FGraph}.

We will now use the trajectory integral matching method\cite{Traject} to compute $\langle \left|u'(\vec{q}, t)\right|^2\rangle$, which determines the presence or absence of smectic order\cite{foot1}.
We restricted this calculation to the case $\Gamma_0<0$, since, as we'll see,  only then is a  stable smectic phase  possible. Performing this standard procedure, 
we obtain
\begin{eqnarray}
\langle \left|u'(\vec{q}, t)\right|^2\rangle
                                            ={\Delta(\ell^*)\over\nu_\perp(\ell^*)}{e^{2\chi\ell^*}\over \left[q_{\perp}^2+{\nu_z(\ell^*)\over\nu_\perp(\ell^*)}q_z ^2\right]},
                                            \label{ucorrtraj}
\end{eqnarray}
where $\ell^*\equiv\ln{\Lambda/q}$, with $\Lambda$  the ultraviolet cutoff. 
For small $q$ ($\ll\Lambda$), $\ell^*\gg 1$; in this limit, we find



\begin{eqnarray}
{\Delta\over\nu_{\perp}}\left(\ell^*\right)&=&\exp\left[-2\chi(\ell^*-\ell_1)+C\int_{\ell_1}^{\ell^*}{d\ell'\over\ell'^{3/2}}\right]\nonumber\\
&&\times
\left({\Delta\over\nu_{\perp}}\left(\ell_1\right)\right),
\label{Delta/nuperpint}
\end{eqnarray}
where $\ell_1$ is some fixed value of the renormalization group``time" $\ell$ at which the large $\ell$ approximations become valid, and  $C$ is a constant that we could express in terms of $\Gamma_0$ and $g_0$ using
Eqns. (\ref{Gammasol}) and (\ref{gsol}), if we cared.

Note that the integral over $\ell'$ in this expression converges as $\ell\to\infty$; hence, 
${\Delta\over\nu_{\perp}}\left(\ell^*\to\infty\right)\to C'e^{-2\chi\ell^*}$, where $C'$ is a finite, non-zero constant. This  convergence  makes the scaling of $u'$ correlations the same as that predicted by the linear theory, as we'll now show.

Using Eq. (\ref{Delta/nuperpint}) in Eq. (\ref{ucorrtraj}) gives
\begin{eqnarray}
\langle  \left|u'(\vec{q}, t)\right|^2\rangle={C'\over \left[q_{\perp}^2+{\nu_z(\ell^*)\over\nu_\perp(\ell^*)}q_z ^2\right]}.
                                            \label{ucorrtraj2}
\end{eqnarray}

Now note that, since $\Gamma\to-1$ as
$\ell\to\infty$,
${\nu_z(\ell^*)\over\nu_\perp(\ell^*)}\to-{\lambda_z(\ell^*)\over\lambda_\perp(\ell^*)}=\left|{\lambda_z^0\over\lambda_\perp^0}\right|$, the last equality following because   the ratio
${\lambda_z(\ell^*)\over\lambda_\perp(\ell^*)}$ does not  renormalize, as can be seen from
the recursion relation (\ref{lambda})  for the $\lambda$'s.
Thus we have, finally,
\begin{eqnarray}
\langle \left|u'(\vec{q}, t)\right|^2\rangle={C'\over \left[q_{\perp}^2+\left|{\lambda_z^0\over\lambda_\perp^0}\right|q_z ^2\right]},
                                            \label{ucorrfinal}
\end{eqnarray}
which clearly scales as $1/q^2$ for all directions of wavevector $\vec{q}$.
This scaling 
implies ``logarithmic roughness" of the smectic layers in $d=2$: that is,
\begin{widetext}
\begin{eqnarray}
\left\langle \left[u'(\vec{r}, t)-u'(\vec{r'}, t)\right]^2\right\rangle={C'\over2\pi}
\sqrt{ \left|{\lambda_\perp^0\over\lambda_z^0}\right|}
\ln{
\left[{\left(r_\perp-r'_\perp\right)^2\over a^2}
+\left|{\lambda_{\perp}^0\over\lambda_z^0}
\right|{
\left(z-z'\right)^2\over a^2}\right]},
                                            \label{ucorr real}
\end{eqnarray}
\end{widetext}
which
implies Eq. (1) via \cite{foot1}
\begin{widetext}
\begin{eqnarray}
 \langle \psi^*(\vec{r},t)\psi(\vec{r}\,',t)\rangle
 \propto\exp{\left[-{1\over 2}q_0^2\left\langle
 \left[u(\vec{r}, t)-u(\vec{r'}, t)\right\rangle\right]^2\right]}\propto\left[{\left(r_\perp-r'_\perp\right)^2\over a^2}
+\left|{\lambda_{\perp}^0\over\lambda_z^0}\right|
{\left(z-z'\right)^2\over a^2}\right]^{-\eta}
 \propto |\vec{r}-\vec{r}\,'|^{-\eta},
 \label{quasi2}
\end{eqnarray}
\end{widetext}
with  $\eta=C'q_0^2\sqrt{\mid\lambda_{\perp}^0/\lambda_z^0\mid}/4\pi$; i.e.,
quasi-long-ranged smectic order\cite{KT}. This is the first real demonstration that the roughness of the anistropic KPZ equation is only logarithmic in $d=2$; earlier arguments 
did not address the possibility that
the integral in equation (\ref{Delta/nuperpint}) could fail to converge as $\ell\rightarrow\infty$,  thereby changing the  scaling  from that predicted by the linearized theory (
as happens in $d=3$ equilibrium smectics\cite{Esmectic}).


When there is no number conservation, as  in the model we are considering here, we
will show later that,
in both $d=2$ and $d=3$,  the spatial Fourier transform of  the number density $\rho$  is given by
\begin{eqnarray}
\langle\left|\rho(\vec{q},t)\right|^2\rangle=C_1 q_z^2\langle \left|u(\vec{q},t)\right|^2\rangle+C_2~,
\label{rhoflucmalt}
\end{eqnarray}
where $C_1$ and $C_2$ are constants.
Since this 
 remains finite as $q\to 0$, 
there are
no giant number fluctuations\cite{GNF}.

Now we discuss the stability of the smectic phase in $d=2$. Doing an anisotropic rescaling $r_\perp''=r_\perp'$, $z''=\sqrt{\nu_{\perp}/\nu_z}z'$, expressing $u'$ in terms of $\theta=2\pi u'/a$, and ignoring the nonlinear terms, we can write Eq. (\ref{AnisoKPZ}) as
\begin{eqnarray}
\partial_{t'} \theta=\nu_{\perp}\nabla''^2 \theta+f',
\label{XY}
\end{eqnarray}
where the statistics of $f'$ are given by
\begin{eqnarray}
 \langle f'(\vec{r}\,'',t')f'(\vec{0},0)\rangle=\kappa\nu_{\perp}\delta\left(\vec{r}\,''\right)\delta(t').
\end{eqnarray}
with $\kappa\equiv\Delta\left(2\pi/a\right)^2/\sqrt{\nu_{\perp}\nu_z}$. Eq. (\ref{XY}) is identical to 
the simplest relaxational equation of motion for an  equilibrium $XY$ model with $\theta$ being the angle of 
the magnetization, 
and $\kappa\nu_{\perp}=2k_BT$. This mapping implies a dislocation unbinding phase transition\cite{Undislocation} in active smectics in $d=2$, when $\kappa=\pi$.
For $\kappa>\pi$, the system is in the ordered fluid phase; for $\kappa<\pi$, the system is in the smectic phase.

The effect of the nonlinear terms can be included simply by replacing the bare $\kappa$ with its (finite)  renormalized value. This implies that the transition occurs when
\begin{eqnarray}
 \lim_{\ell\to\infty}\kappa(\ell)=\pi.\label{Critical}
\end{eqnarray}
Note that Eq. (\ref{Critical}) is valid only if the period of the rescaled system on $u'$ is kept fixed at $a$, which requires no $u'$ rescaling (i.e.,  $\chi=0$). Using our earlier solution to the recursion relations it is straightforward to show that
\begin{eqnarray}
  \lim_{\ell\to\infty}\kappa(\ell)\to -{\left(1-\Gamma_0\right)^2\over 4\Gamma_0}\kappa_0.
\end{eqnarray}
Using this result in Eq. (\ref{Critical}) we obtain the phase boundary in terms of $\kappa_0$ and $\Gamma_0$ for $\Gamma_0<0$
\begin{eqnarray}
 \kappa_0=-{4\pi\Gamma_0\over\left(1-\Gamma_0\right)^2}.\label{BareCritical}
\end{eqnarray}
This phase diagram in the $\kappa_0$-$\Gamma_0$ parameter space is illustrated in Fig. \ref{fig: ActivePhase}.

Now we turn to the case where the number of particles is conserved. In this case the fluctuation $\delta\rho\equiv\rho-\rho_0$ of the density $\rho$ about its mean value $\rho_0$ becomes another important hydrodynamical variable. Number conservation implies $\partial_t\delta\rho=-\vec{\nabla}\cdot\vec{j}$, where $\vec{j}$ is the number density current. Based on  symmetry arguments, a gradient expansion of $\vec{j}$, keeping only ``relevant" terms 
is given by $\vec{j}=\vec{j}_L+\vec{j}_{NL}$, where the linear piece
\begin{eqnarray}
 \vec{j}_L&=&-[j_0+v_\rho\delta\rho+D_z\partial_z\delta\rho+v^{z}_{\rho u}\partial_zu
 \nonumber\\&&+\left((c_\perp-w)\nabla^2_\perp+c_z\partial_z^2\right)u]\hat{z}\nonumber\\
 &&-D_{\perp}\vec{\nabla}_{\perp}\delta\rho-v^\perp_{\rho u}\vec{\nabla}_{\perp}u-w\vec{\nabla}_\perp\partial_zu-\vec{f}_\rho~,
 \label{linear Current}\end{eqnarray}
where $\vec{f}_\rho$ is a Gaussian noise with  statistics
\begin{eqnarray}
 \langle f_{\rho i}(\vec{r},t)f_{\rho j}(\vec{0},0)\rangle=\left(\Delta_z\delta_{ij}^z+\Delta_{\perp}\delta_{ij}^{\perp}\right)\delta(\vec{r})\delta(t)~,
\end{eqnarray}
while the non-linear piece is given by
\begin{eqnarray}
 \vec{j}_{NL}&=&-[ \lambda_{\perp\rho} |\vec{\nabla}_{\perp}u|^2+\lambda_{z\rho} \left(\partial_{z} u\right)^2+v_\rho\delta\rho\partial_zu
+g_\rho\delta\rho^2]\hat{z} \nonumber\\&&+g_u\partial_zu\vec{\nabla}_\perp u+v_\rho\delta\rho\vec{\nabla}_\perp u~.
 \label{non-linear Current}\end{eqnarray}
Similar symmetry arguments and gradient expansions give the equation of motion  for $u$, again keeping all relevant terms,
\begin{eqnarray}
\partial_{t} u&=&v_0+v_u\partial_z u+v_{u\rho}\delta\rho+\nu_z\partial_{z}^2 u + \nu_{\perp}\partial_{\perp}^2 u + \nu_{\rho}\partial_z\delta\rho\nonumber\\
 &&+\lambda_{\perp} |\vec{\nabla}_{\perp}u|^2+\lambda_z \left(\partial_{z} u\right)^2+g\delta\rho^2\nonumber\\
 &&+g_c\delta\rho\partial_zu+ f_u,
\label{EOM_u}
\end{eqnarray}
where the noise $f_u$ has the same  statistics as $f$ in Eq. (\ref{AnisoKPZ}).

If we neglect the non-linear terms in $\vec{j}$ and (\ref{EOM_u}), a straightforward calculation shows that $\langle \left|u(\vec{q},t)\right|^2\rangle\sim 1/q^2$, which implies quasi-long-ranged smectic order in $d=2$ and long-ranged order in $d=3$. We also find that $\langle \left|\delta\rho(\vec{q},t)\right|^2\rangle$ goes to a finite value as $q\to 0$, which implies no giant number fluctuations in either $d=2$ or $d=3$.

Simple power counting shows that the nonlinear terms in $\vec{j}$ and (\ref{EOM_u}) are irrelevant in $d=3$, in the RG sense. Hence, these linear results should apply in $d=3$, at least in systems with sufficiently small non-linearities\cite{strong coupling KPZ footnote}.
In $d=2$, similar power counting shows that all of the  nonlinear terms in $\vec{j}$ and (\ref{EOM_u}) become marginal, and, hence, could potentially change the behavior at long wavelengths.  


Supplementing the continuity equation for $\delta\rho$
with a source term to reflect the tendency of birth and death to restore the local population density to its equilibrium value, and an additional, non-number-conserving noise reflecting statistical fluctuations in the local birth and death rate,  as has been done\cite{malt} for flocks with polar orientational order, enables us to analyze number fluctuations in
the Malthusian case as well. Dropping irrelevant terms, we obtain the equation of motion:

\begin{eqnarray}
 \partial_{t} \delta\rho=\alpha\partial_z u+
 v_{\rho u}^\perp\nabla_{\perp}^2 u +v^z_{\rho u}\partial_{z}^2 u -\delta\rho/\tau +f_{b-d}~,
 \label{EOM_rho malt}
\end{eqnarray}
where $\tau$ is the characteristic relaxation time for density fluctuations to relax away due to birth and death\cite{malt}, the $\alpha$ term reflects the fact that symmetry allows the local birth and death rate to depend on the local layer spacing, and $f_{b-d}$ is the aforementioned noise in the birth and death rate. We take $f_{b-d}$ to be zero mean Gaussian white noise, with statistics  $\left<f_{b-d}(\vec{r}, t)f_{b-d}(\vec{r}~', t')\right>=2\Delta_{b-d}\delta^d(\vec{r}-\vec{r}~')\delta(t- t')$.

Fourier transforming equation (\ref{EOM_rho malt}) in space, and solving the resultant linear stochastic ODE for the correlations of $\delta\rho$ gives\cite{rhoFT}, to leading order in $q$, equation (\ref{rhoflucmalt}), with $C_1=\tau^2\alpha^2$ and $C_2=\tau\Delta_{b-d}$.

In conclusion, we have developed the  hydrodynamic theory  of apolar active smectic in both $d=2$ and $d=3$. We considered both the  case in which the number of the particles is  conserved, and that in which it is not.  We have made various experimental predictions which can be tested. For the number conserving case an RG analysis of the nonlinear terms remains to be done.

We thank P.  Romanczuk for invaluable discussions, and  the MPIPKS, Dresden, where those discussions took place, for their support (financial and otherwise) and hospitality.  JT thanks the U.S.  National Science Foundation for their financial support
through awards \# EF-1137815 and 1006171; LC acknowledges support by the National Science Foundation of China (under
Grant No. 11004241) and the Fundamental Research
Funds for the Central Universities (under Grant No.
2010LKWL09). LC also thanks the China Scholarship Fund for supporting his visit to the University of Oregon, where a portion of this work was done. He is also grateful to the University of Oregon's Physics Department and Institute for Theoretical Science for their hospitality.


\begin{thebibliography}{99}
\bibitem{amplitude flucts nonhydro}In principle, the amplitude $\left|\psi_0\right|$
can also fluctuate; however, such fluctuations are non-hydrodynamic; that is, their relaxation rate remains non-zero as $q\rightarrow 0$. Hence, these fluctuations can be ignored in a hydrodynamic treatment.
\bibitem{foot1}
The presence or absence of smectic order depends on the behavior of the equal-time correlation function $\langle [u(\vec{r}, t)-u(\vec{r}\,', t)]^2\rangle$. In this problem $\langle [u(\vec{r}, t)-u(\vec{r}\,', t)]^2\rangle=\langle [u'(\vec{r}, t)-u'(\vec{r}\,', t)]^2\rangle$ by Eq. (\ref{uprime}). To calculate  $\langle [u'(\vec{r}, t)-u'(\vec{r}\,', t)]^2\rangle$, we only need $\langle \left|u\,'(\vec{q},t)\right|^2\rangle$.
\end{thebibliography}
\end{document}